\documentclass[useAMS,usenatbib]{mn2e}

\usepackage{graphicx,psfrag}

\title[What is galaxy environment?]{Measures of Galaxy Environment --\\ I. What is `Environment'?}
\author[Muldrew et al.]{Stuart I. Muldrew$^{1}$\thanks{E-mail: ppxsm2@nottingham.ac.uk}, Darren J. Croton$^{2}$, Ramin A. Skibba$^{3}$, Frazer R. Pearce$^{1}$, \newauthor
Hong Bae Ann$^{4}$, Ivan K. Baldry$^{5}$, Sarah Brough$^{6}$, Yun-Young Choi$^{7}$, \newauthor
Christopher J. Conselice$^{1}$, Nicolas B. Cowan$^{8}$, Anna Gallazzi$^{9}$, Meghan E. Gray$^{1}$, \newauthor
Ruth Gr\"{u}tzbauch$^{1}$, I-Hui Li$^{2}$, Changbom Park$^{10}$, Sergey V. Pilipenko$^{11}$, \newauthor
Bret J. Podgorzec$^{1}$, Aaron S.G. Robotham$^{12}$, David J. Wilman$^{13}$, Xiaohu Yang$^{14}$, \newauthor
Youcai Zhang$^{14}$ and Stefano Zibetti$^{9}$\\
$^{1}$School of Physics and Astronomy, University of Nottingham, Nottingham, NG7 2RD, UK\\
$^{2}$Centre for Astrophysics \& Supercomputing, Swinburne University of Technology, P.O. Box 218, Hawthorn, VIC 3122, Australia\\
$^{3}$Steward Observatory, University of Arizona, 933 N. Cherry Avenue, Tucson, AZ 85721, USA\\
$^{4}$Division of Science Education, Pusan National University, Busan 609-735, South Korea\\
$^{5}$Astrophysics Research Institute, Liverpool John Moores University, Twelve Quays House, Egerton Wharf, Birkenhead, CH41 1LD, UK\\
$^{6}$Australian Astronomical Observatory, PO Box 296, Epping, NSW 1710, Australia\\
$^{7}$Department of Astronomy and Space Science, Kyung Hee University, Gyeonggi 446-701, South Korea\\
$^{8}$Northwestern University, Dearborn Observatory, 2131 Tech Drive, Evanston, IL 60208, USA\\
$^{9}$Dark Cosmology Centre, Niels Bohr Institute, University of Copenhagen, Juliane Maries Vej 30, DK-2100 Copenhagen, Denmark\\
$^{10}$Korea Institute for Advanced Study, 87 Hoegiro, Dongdaemun-Gu, Seoul 130-722, South Korea\\
$^{11}$Astro Space Center, Lebedev Physical Institute, Russian Academy of Sciences, 117997 Moscow, Russia\\
$^{12}$School of Physics and Astronomy, University of St Andrews, North Haugh, St Andrews, Fife, KY16 9SS, UK\\
$^{13}$Max-Planck-Institut f\"{u}r Extraterrestrische Physik, Giessenbachstra\ss e, D-85748 Garching, Germany\\
$^{14}$Key Laboratory for Research in Galaxies and Cosmology, Shanghai Astronomical Observatory (the Partner Group of MPA),\\
\hspace{4pt} Nandan Road 80, Shanghai 200030, China}
\begin{document}

\date{Accepted 2011 September 28. Received 2011 September 27; in original form 2011 August 23}

\pagerange{1--14} \pubyear{2012}

\maketitle

\label{firstpage}

\begin{abstract}
The influence of a galaxy's environment on its evolution has been studied and compared extensively in the literature, although differing techniques are often used to define environment. Most methods fall into two broad groups: those that use nearest neighbours to probe the underlying density field and those that use fixed apertures.  The differences between the two inhibit a clean comparison between analyses and leave open the possibility that, even with the same data, different properties are actually being measured. In this work we apply twenty published environment definitions to a common mock galaxy catalogue constrained to look like the local Universe.  We find that nearest neighbour-based measures best probe the internal densities of high-mass haloes, while at low masses the inter-halo separation dominates and acts to smooth out local density variations.  The resulting correlation also shows that nearest neighbour galaxy environment is largely independent of dark matter halo mass. Conversely, aperture-based methods that probe super-halo scales accurately identify high-density regions corresponding to high mass haloes.  Both methods show how galaxies in dense environments tend to be redder, with the exception of the largest apertures, but these are the strongest at recovering the background dark matter environment.  We also warn against using photometric redshifts to define environment in all but the densest regions.  When considering environment there are two regimes: the `local environment' internal to a halo best measured with nearest neighbour and `large-scale environment' external to a halo best measured with apertures.  This leads to the conclusion that there is no universal environment measure and the most suitable method depends on the scale being probed.
\end{abstract}


\section{Introduction}

In the paradigm of hierarchical structure formation, the evolution of the primordial density field acting under gravitational instability drives dark matter to cluster and collapse into virialised objects (haloes). Such haloes provide the potential wells into which baryons fall and galaxies subsequently form \citep{White78}. Haloes, galaxies and their environments also interact and merge as structure formation unfolds with time. It therefore follows that the properties of a galaxy should be correlated with the properties of its host halo, and that a galaxy's environment, its host halo's environment, and the dark matter density field are all related in some measurable way.

Such galaxy/halo/dark matter correlations with environment have lead to a variety of work examining the environmental dependence of the physics of galaxy formation, both theoretical and observational. Measurements of the galaxy two-point correlation function and halo occupation distribution function (HOD) have shown that more massive, brighter, redder, and passive early-type galaxies tend to be more strongly clustered and hence presumably located in denser environments, while the reverse is true for galaxies that have lower mass, are fainter, bluer and star forming \citep[e.g.][]{Norberg02, Zehavi05, Sheth06, Li06, Tinker08, Ellison09, Skibba09b, Skibba09c, delaTorre11}.

A more direct probe of the influence of environment is the local density field of neighbouring galaxies around each galaxy (defined in various ways).  These techniques are better suited to analysing targeted halo and galaxy environment correlations and have proven valuable in the current era of large galaxy survey data sets, where galaxy catalogues can be simultaneously `sliced' in multiple orthogonal directions to isolate the dependence of specific galaxy properties on environment \citep[e.g.][]{Kauffmann04, Blanton05, Croton05, Cooper06, Baldry06, Park07, Elbaz07, Ball08, Cowan08, OMill08, Tasca09, Ellison09}.

In undertaking any such analysis the choice of environmental indicator is important and no one standard has yet emerged. Many of the above cited papers involve disparate selection criteria, research methods and goals, making direct comparisons between them difficult. The definition of environment can vary from two-point clustering and marked clustering statistics, to the number or luminosity density within a fixed spherical or cylindrical aperture, to the measured density enclosed by the $n$-th nearest neighbour. A further complication is that these methodologies can be performed in either two (projected) or three (redshift space) dimensions. As a consequence, some analyses have yielded irreconcilable results.

All methods that attempt to quantify the environment around a galaxy require some parameter choices. Those that involve a spherical or cylindrical aperture must first choose a fixed smoothing scale within which to measure the local galaxy over- or under-density. On the other hand, when environment measures involve the $n$-th nearest neighbour, the choice of $n$ instead becomes important. Once $n$ is fixed, this statistic adapts its scale to keep the signal-to-noise constant. But how should one interpret a statistic that combines the physical processes from widely disparate scales across one smoothly varying curve? And how should this be compared with statistics that instead fix the scale along the same curve?

Further complicating comparison are the selection criteria of a dataset itself, its geometry and volume, and the redshift and magnitude uncertainties of the galaxies in it.  In short, the measurement of `environment' used in various studies can be completely different, and environmental correlations should be interpreted and compared with caution.  Some environment measures can have advantages and disadvantages for particular research goals.  A number of authors have tested and compared a few environment measures \citep[e.g.][]{Cooper05, Wolf09, Gallazzi09, Kovac10, Wilman10, Haas11}.  In general, while the environment measures are correlated, they often exhibit considerable scatter between them.

The primary goal of this project is to compare a variety of published environment measures using a single well constrained data set.  For this purpose, we take a dark matter halo catalogue and construct a mock galaxy catalogue designed to have approximately the same global statistical properties as the Sloan Digital Sky Survey \citep[SDSS;][]{York00} main galaxy sample.  We then are able to compare the galaxy environment measures to halo mass, dark matter density, and to each other.  We also attempt to answer some important questions, such as: Do the different environment methodologies break nicely into different groups that optimally sample the underlying density field in particular ways?  Do the statistics of various galaxy properties change dramatically in different environment bins measured in different ways?  Can we find a more fundamental definition of environment that is measurable observationally?

This paper is organised as follows: In Section \ref{sec:mod} we outline the mock galaxy catalogue that was generated, constrained by the SDSS, and used to study environment measures.  In Section \ref{sec:meas} we review the range of environment measures available in the literature that are used as part of this study.  Having established the method, Section \ref{sec:res} explores how the different measures relate to the dark matter halo mass, galaxy colour and large-scale dark matter environment for each galaxy.  We also explore how the measures relate to each other for an individual galaxy.  Finally in Section \ref{sec:discussion} we discuss and summarise our findings.

\section{Galaxy and Halo Catalogues}
\label{sec:mod}

\subsection{The Millennium Dark Matter Simulation}
\label{sec:MS}

We begin with the Millennium Simulation \citep[][]{Springel05} which is a large $N$-body simulation of dark matter structure in a cosmological volume.  The Millennium Simulation uses the \textsc{gadget} Tree-PM code \citep{Springel05a} to trace the evolution of 10 billion dark matter particles across cosmic time in a cubic box of $500 h^{-1} {\rm Mpc}$ on a side, with a halo mass resolution of $\sim5\times10^{10}h^{-1}M_\odot$.  It adopts the concordance $\Lambda$CDM cosmological parameters, chosen to agree with a combined analysis of the Two-Degree Field Galaxy Redshift Survey \citep[2dFGRS;][]{Colless01} and the first-year Wilkinson Microwave Anisotropy Probe data \citep[WMAP;][]{Spergel03}: $\Omega_0=0.25$, $\Omega_\Lambda=0.75$, $h=0.73$, $n=1$, and $\sigma_8=0.9$. 

The haloes are found by a two-step procedure.  First, all collapsed haloes with at least 20 particles are identified using a standard friends-of-friends group-finder with linking parameter $b=0.2$.  Then, post-processing with the substructure algorithm \textsc{subfind} \citep{Springel01} subdivides each friends-of-friends halo into a set of self-bound subhaloes.  We note that comparable halo properties are found using other structure finders \citep[see][]{Knebe11}.

\subsection{Embedding Galaxies in Haloes}
\label{sec:HOD}

\begin{figure}
\psfrag{Halo Mass}[][][1][0]{$\rm log(M_{Halo}/M_{\odot})$}
\psfrag{Galaxies}[][][1][0]{$\langle N_\mathrm{gal}|M,L_\mathrm{min}\rangle$}
\psfrag{a}[l][][1][0]{$M_r-5{\rm log}(h)\leq-19$}
\psfrag{b}[l][][1][0]{$M_r-5{\rm log}(h)\leq-20$}
\includegraphics[width=82mm]{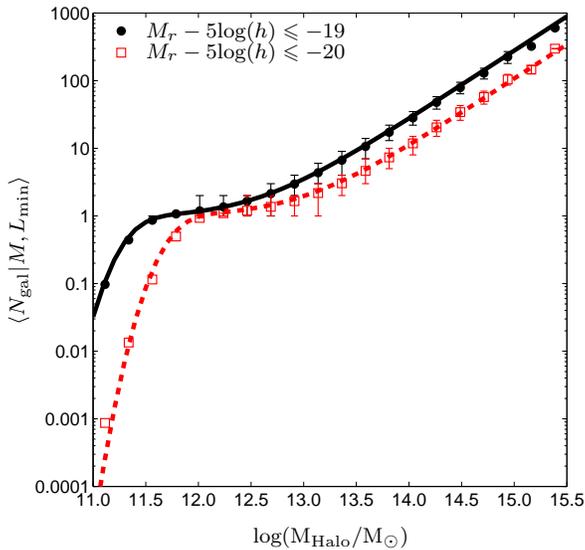}
\caption{The mean number of galaxies above a given luminosity present in dark matter haloes of different mass.  Error bars denote the 16th and 84th percentiles and are plotted for haloes that on average host at least one galaxy. Lines represent the input model and correspond to Equation \ref{fit}. [Note this Figure differs from the published version as this one contains all haloes, while the published version contains only haloes hosting at least one galaxy.]}
\label{fig:pop}
\end{figure}

\begin{figure*}
  \includegraphics[width=0.497\hsize]{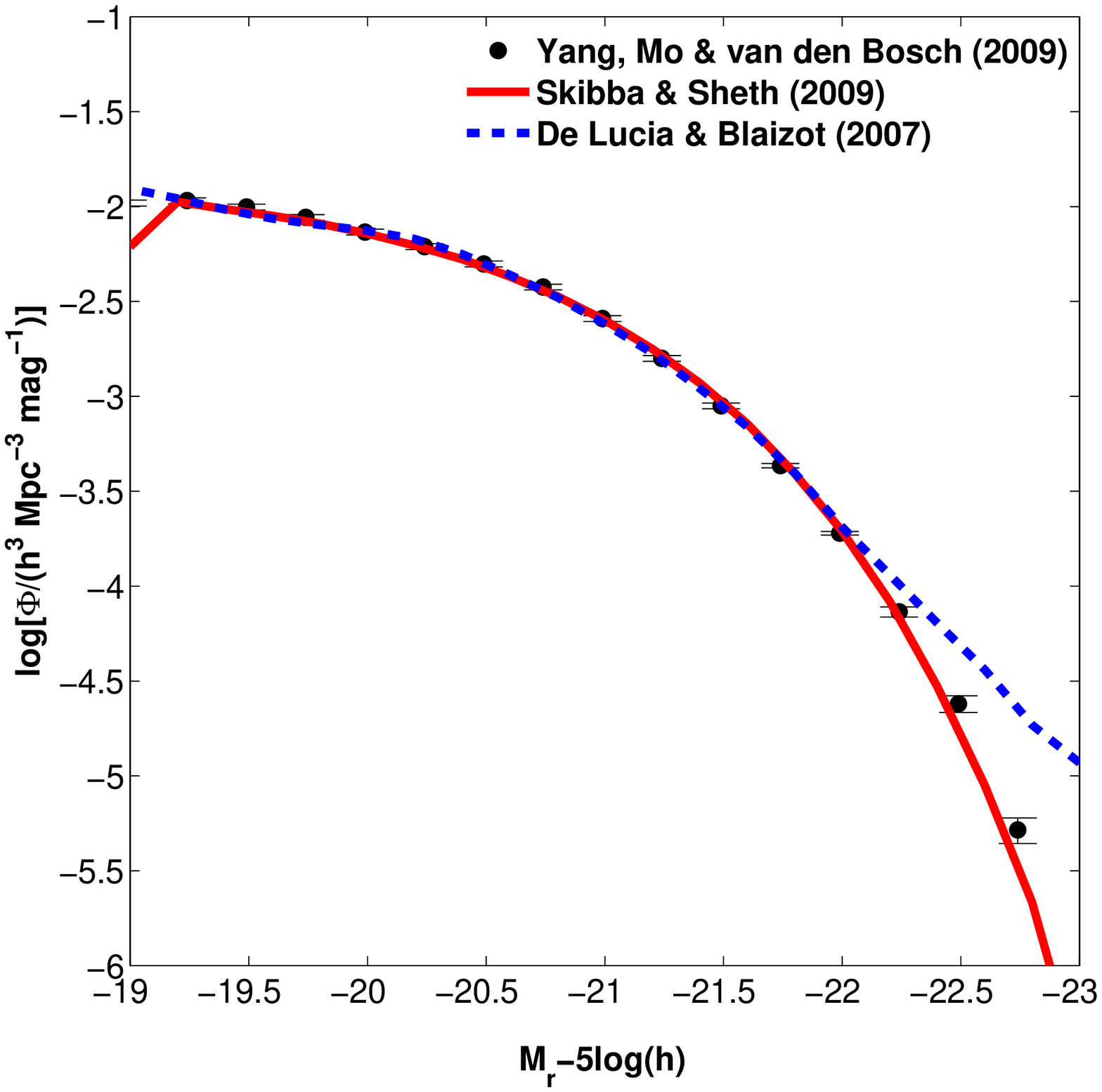} 
  \includegraphics[width=0.497\hsize]{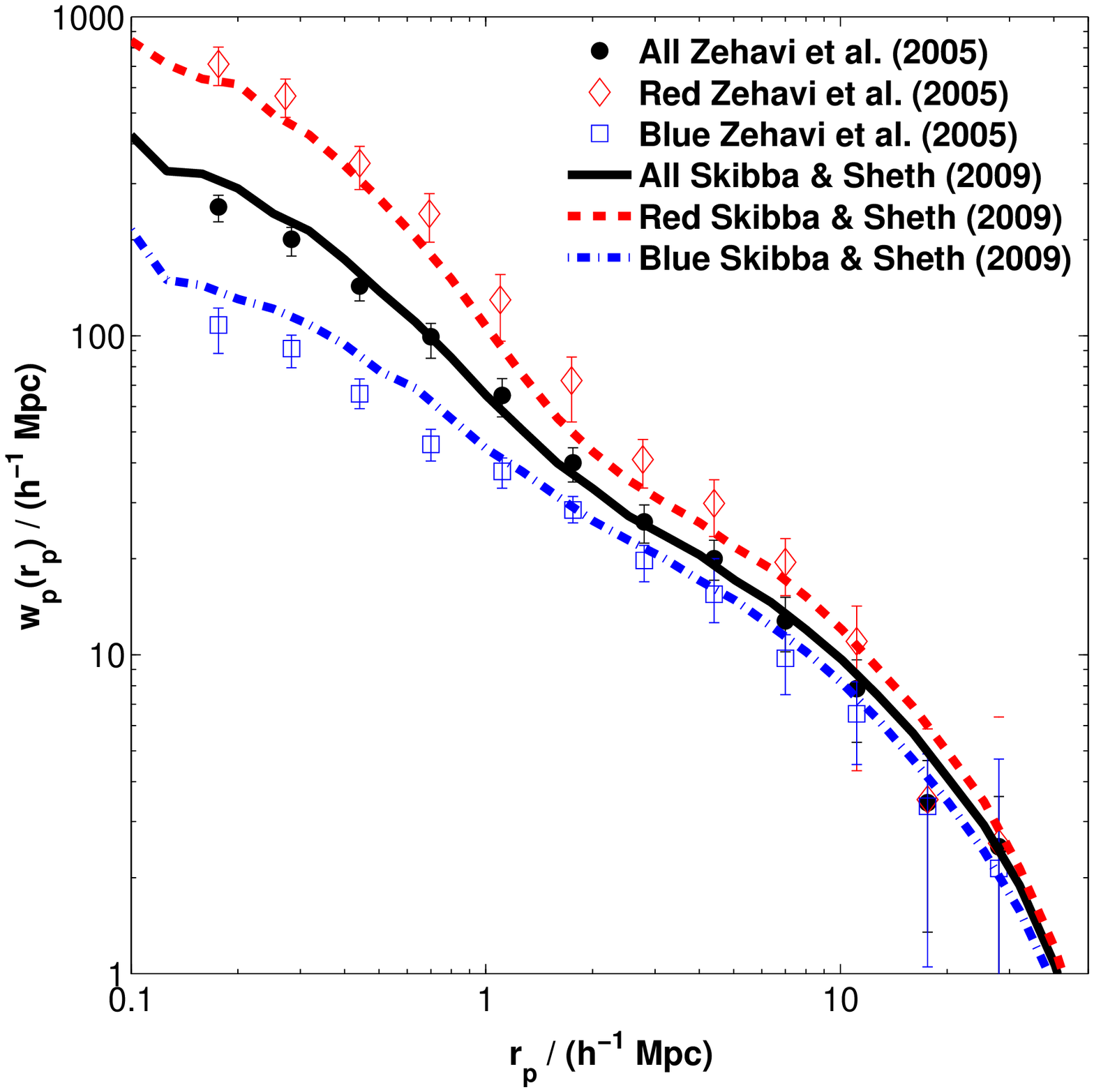} 
  \caption{Left: the $r$-band luminosity function for the mock galaxy catalogue 
           created using the HOD of \citet{Skibba09b} (red line) compared with 
           that of the semi-analytic \citet{DeLucia07} model (blue line) and the 
           SDSS observed values \citep{Yang09} (black points with errors).
           Right: Two-point correlation function of all, red, and blue galaxies in the mock 
           catalogue (lines), compared with the equivalent observed results in the SDSS 
           from \citet{Zehavi05} (points with errors).}
\label{fig:lumfunc}
\end{figure*}

From the Millennium Simulation halo merger tree at $z=0$, we construct a mock galaxy catalogue using the halo occupation method described in \citet[][ hereafter S06]{Skibba06} and \citet[][ hereafter SS09]{Skibba09b}; we refer the reader to these papers for details.  Other halo-model descriptions of galaxy clustering---conditional luminosity functions \citep[e.g.][]{Yang03} and subhalo abundance matching \citep[e.g.][]{Kravtsov04}---would produce similar mock catalogues, although an advantage of the SS09 approach is that it includes a strongly constrained model of galaxy colours.  S06 describes how the luminosities and real-space and redshift-space galaxy positions are modelled.

Our model distinguishes between the `central' galaxy in a halo and all the other galaxies (`satellites').  We assume that central galaxies have the same positions and velocities as the haloes in the dark matter simulation.  In other words, central galaxies are at the centre of the haloes, and the satellites are located around them.  An important assumption in the model is that all galaxy properties---their numbers, spatial distributions, velocities, luminosities, and colours---are determined by halo mass alone.  These galaxy properties are constrained by SDSS observations, including the luminosity function \citep{Blanton03}, luminosity-dependent two-point clustering \citep{Zehavi05, Skibba06, Zheng07}, and the colour-magnitude distribution and colour-dependent clustering \citep{Skibba09a}.  Note that the clustering constraints result in a mock catalogue that approximately reproduces the observed environmental dependence of luminosity and colour, on scales of $100\,h^{-1} {\rm kpc}$ to $30\,h^{-1} {\rm Mpc}$.

The number of satellite galaxies in the model follows a Poisson distribution with a mean value that increases with halo mass.  The satellites are distributed around the halo centre so that they follow a \citet{Navarro96} profile with the mass-concentration relation from \citet{Maccio08}.  We assign redshift-space coordinates to the mock galaxies assuming that a galaxy's velocity is given by the sum of the velocity of its parent halo plus a virial motion contribution that is drawn from a Maxwell-Boltzmann distribution with dispersion that depends on halo mass (S06).

We specify a minimum $r$-band luminosity for the galaxies in the catalogue, $M_r-5{\rm log}(h)=-19$, to stay well above the resolution limit of the Millennium Simulation, avoiding any issues of completeness that may bias our results.  We generate luminosities for the central galaxies, while accounting for the stochasticity between their luminosities and host halo mass, and then we generate the satellite luminosities so that the observed luminosity distribution is reproduced for $M_r-5{\rm log}(h)\leq-19$ (S06). 

We model the observed $g-r$ colour distribution at a given luminosity as the sum of two Gaussian components, commonly referred to as the `blue cloud' and `red sequence'.  Our colour model has five constraints as a function of luminosity: the mean and scatter of the red sequence, mean and scatter of the blue cloud, and the blue fraction.  We assume that the colour distribution at fixed luminosity is approximately independent of halo mass, and that the satellite colour distribution varies such that its mean increases with luminosity (i.e., the satellite red fraction increases with luminosity in a particular way).  These two assumptions are tested and verified with galaxy group catalogues in \citet{Skibba09a}.

This procedure produces a mock galaxy catalogue containing 1.84 million galaxies, of which 29 percent are satellites.  Galaxies occupy haloes with masses ranging from $10^{11}$ to $10^{15.3}h^{-1} M_\odot$.  We also construct a mock light cone from the catalogue by selecting galaxies that are within a radial distance of $500\,h^{-1} {\rm Mpc}$ from one corner of the box.  This gives an opening angle of $90\times90$ degrees and a depth of $500\,h^{-1} {\rm Mpc}$, for which right ascension and declinations are determined.  The analysis in Section~\ref{sec:res} is carried out using a sample of galaxies that are common to both the box and the cone and are chosen so not to be affected by edges.  Figure~\ref{fig:pop} shows the mean number of galaxies as a function of halo mass, for two luminosity thresholds ($L_\mathrm{min}$).  By construction, the number of galaxies consists of the number of central galaxies plus the number of satellites, such that 
\begin{equation}
 \langle N_\mathrm{gal}|M,L_\mathrm{min}\rangle \,=\,
    \langle N_{\mathrm {cen}}|M,L_\mathrm{min}\rangle\,
    \Bigl[1 + \langle N_{\mathrm {sat}}|M,L_\mathrm{min}\rangle\Bigr]
\label{fit}
\end{equation}
\noindent where,
\begin{equation}
  \langle N_\mathrm{cen}|M\rangle \,=\, \frac{1}{2}\Biggl[1\,+\,\mathrm{erf}\Biggl(\frac{\mathrm{log}(M/M_\mathrm{min})}{\sigma_{\mathrm{log}M}}\Biggr)\Biggr]
\label{NcenM}
\end{equation}
\noindent and
\begin{equation}
  \langle N_\mathrm{sat}|M\rangle = 
    \Biggl(\frac{M-M_0}{M_1^{ ' }}\Biggr)^\alpha .
\label{NsatM}
\end{equation}
\noindent (See Appendix A2 of SS09 for details).  All of the free parameters depend on luminosity.  The slope of the power law, $\alpha$, is nearly unity.  One may define a parameter $M_1$, which is equal to or slightly larger than $M_1^{'}$ \citep{Zheng07}, and is proportional to the minimum halo mass: $M_1\approx 20\,M_\mathrm{min}$.  This determines the mass above which haloes typically host at least one satellite galaxy.  Therefore, since for $M_r\leq-19$ the minimum halo mass is $\approx10^{11.5}h^{-1}M_\odot$, the mean number of galaxies rises rapidly like a linear power law at masses larger than twenty times this value, or $\approx10^{12.8}h^{-1}M_\odot$, as seen in Figure~\ref{fig:pop}.

At the high halo mass end, galaxy number shows a near linear relationship with dark matter halo mass, which occurs by construction in the halo occupation model.  This implies that the number of galaxies per unit dark matter mass is constant, or put another way, each galaxy contributes the same mass of dark matter to the cluster.  This is in agreement with the findings of \citet{Poggianti10} and to some degree is the natural consequence of a structure built hierarchically.  This also agrees with \citet{Blanton07} who find that galaxy distributions are only affected by the host dark matter halo, and not by the surrounding density field, for the SDSS galaxy group catalogue.

The $r-$band luminosity function of galaxies in the mock catalogue is shown in the left-hand panel of Figure~\ref{fig:lumfunc} and is compared to both the observed SDSS luminosity function \citep{Yang09} and a popular semi-analytic galaxy formation model \citep{DeLucia07}. In the right-hand panel of Figure~\ref{fig:lumfunc} we show the mock two-point correlation functions of all, red, and blue galaxies and compare them with the equivalent SDSS measurements of \citet{Zehavi05}. Note that the colour-dependent two-point function measured by \citet{Zehavi11} is slightly different from that constrained in the mock, likely due to the presence of the Sloan Great Wall in the real data, an unusually massive supercluster at $z\sim0.08$.

We have made the mock galaxy catalogue as realistic as possible, and although the catalogue reproduces the observed environmental dependence of luminosity and colour, there are nonetheless a few limitations to the model.  For example, we have assumed virialised (dynamically relaxed) dark matter haloes even though some haloes are not, such as those having recently experienced a merger \citep[e.g.][]{Maccio07}.  We have also assumed that central galaxies are always the brightest galaxy in a halo and lie at the centre of their potential well, although in a nonzero fraction of haloes, especially massive haloes, this assumption is not valid \citep{Skibba11}.  Finally, we force satellite galaxy properties to depend only on halo mass, not on halo-centric position, although there is evidence of such a dependence at fixed mass \citep[e.g.][]{Bosch08, Hansen09}.
While our mock galaxy catalogue resembles a spectroscopic catalogue, some environment measures used in the literature are based on photometric data \citep[e.g.][]{Gallazzi09}; for tests with such measures one can add scatter to the redshifted mock galaxy positions, for example.

\begin{table*}
 \begin{tabular}{ c | c | c }
   \hline
   Num. & Method & Author \\ \hline
     & \textbf{Neighbours} & \\
   1 & 3rd Nearest Neighbour & Muldrew\\
   2 & Projected Voronoi & Podgorzec \& Gray \\
   3 & Mean 4th \& 5th  Nearest Neighbour & Baldry$^1$ \\
   4 & 5 Neighbour Cylinder & Li$^2$ \\
   5 & 7th Projected Nearest Neighbour & Ann \\
   6 & 10 Neighbour Bayesian Metric & Cowan$^3$ \\
   7 & 20 Neighbour Smooth Density & Choi \& Park$^4$ \\
   8 & 64 Neighbour Smooth Density & Pearce \\
   \\
     & \textbf{Aperture} \\
   9 & $1\,h^{-1} {\rm Mpc}$ ($\pm1000\,\rm km\,s^{-1}$) & Gr\"{u}tzbauch \& Conselice$^5$ \\
   10 & $2\,h^{-1} {\rm Mpc}$ ($\pm500\,\rm km\,s^{-1}$) & Gallazzi$^6$ \\
   11 & $2\,h^{-1} {\rm Mpc}$ ($\pm1000\,\rm km\,s^{-1}$) & Gr\"{u}tzbauch \& Conselice \\
   12 & $2\,h^{-1} {\rm Mpc}$ ($\pm6000\,\rm km\,s^{-1}$) & Gallazzi$^6$ \\
   13 & $5\,h^{-1} {\rm Mpc}$ ($\pm1000\,\rm km\,s^{-1}$) & Gr\"{u}tzbauch \& Conselice \\
   14 & $8\,h^{-1} {\rm Mpc}$ Spherical & Croton$^7$ \\
  \\
      & \textbf{Annulus} \\
   15 & $0.5-1.0\,h^{-1} {\rm Mpc}$ ($\pm1000\,\rm km\,s^{-1}$) & Wilman \& Zibetti$^8$ \\
   16 & $0.5-2.0\,h^{-1} {\rm Mpc}$ ($\pm1000\,\rm km\,s^{-1}$) & Wilman \& Zibetti$^8$ \\
   17 & $0.5-3.0\,h^{-1} {\rm Mpc}$ ($\pm1000\,\rm km\,s^{-1}$) & Wilman \& Zibetti$^8$ \\
   18 & $1.0-2.0\,h^{-1} {\rm Mpc}$ ($\pm1000\,\rm km\,s^{-1}$) & Wilman \& Zibetti$^8$ \\
   19 & $1.0-3.0\,h^{-1} {\rm Mpc}$ ($\pm1000\,\rm km\,s^{-1}$) & Wilman \& Zibetti$^8$ \\
   20 & $2.0-3.0\,h^{-1} {\rm Mpc}$ ($\pm1000\,\rm km\,s^{-1}$) & Wilman \& Zibetti$^8$ \\
\end{tabular}
\caption{List of environment measures used in this study and the authors who implemented them, including references where applicable.  See Section \ref{sec:meas} for further details.  References: 1: \citet{Baldry06}, 2: \citet{Li11}, 3: \citet{Cowan08}, 4: \citet{Park07}, 5: \citet{Gruetzbauch11}, 6: \citet{Gallazzi09}, 7: \citet{Croton05} and 8: \citet{Wilman10}.}
\label{tab:comp}
\end{table*}

\section{Environmental Measures}
\label{sec:meas}

There are many different methods of measuring galaxy environment available in the literature.  Most of these can be categorised into two broad groups: those which use neighbour finding and those that use a fixed aperture.  An overview of the methods used in this work are presented in the following subsections and summarised in Table~\ref{tab:comp} along with the authors who implemented them.

\subsection{Nearest Neighbour Environment Measures}
\label{sec:neigh}

The principle of nearest neighbour finding is that galaxies with closer neighbours are in denser environments. To create a standard measure for this, a value of $n$ is chosen that specifies the number of neighbours around the point of interest. In its simplest form, the projected surface density of galaxies, $\sigma_{n}$, can then be defined as
\begin{equation}
\sigma_{n}=\frac{n}{\pi r_{n}^2} ~,
\label{eqn:nn2d}
\end{equation}
\noindent where $n$ is the number of neighbours within the projected distance $r_{n}$, the radius to the $n$-th nearest neighbour. 
One disadvantage of quantifying environment using projected statistics is that two galaxies can appear close together when they are in fact just a chance alignment and are actually separated by a larger distance in the third dimension.  While there is no simple way to overcome this observationally, one can adopt a velocity cut about each galaxy, typically of order $\pm$1000 km $\rm s^{-1}$, to minimise the number of such alignments. 
 
For data where a third dimension has been measured for each galaxy (e.g. redshift), the denominator of Equation~\ref{eqn:nn2d} is replaced by the enclosed volume:
\begin{equation}
\Sigma_{n}=\frac{n}{(4/3) \pi r_{n}^3} ~.
\end{equation}
\noindent When using three dimensions careful consideration of redshift distortions are needed and this often leads to two dimensional projected distances often being used.  The nearest neighbour estimator was recently applied to the Galaxy and Mass Assembly catalogue \citep[GAMA;][]{Driver11} by \citet{Brough11} using the distance to the first nearest neighbour above a given luminosity, although typically 3-10 neighbours are used.

Variations on the $n$-th nearest neighbour approach have been proposed in an attempt to improve the robustness of statistic as a measure of local density.  One such method used by \citet{Baldry06} was to take the average of two different neighbour densities, in their case the 4th and 5th nearest neighbour projected surface densities.  An alternative proposed by \citet{Cowan08} was to use the distance to every neighbour up to the tenth instead of just the distance to the tenth to calculate the density.  They adopted a Bayesian metric such that
\begin{equation}
\phi=C\frac{1}{\sum_{i=1}^{10}d_i^3} ~,
\end{equation}
\noindent where $C=11.48$ is empirically determined so that the mean of $\phi$ matches the number density when the density is estimated on a regular grid for a uniform field, and $d_i$ is the distance to neighbour $i$.

One can also use numerical simulations to guide the nearest neighbour calibration. Calculating densities using neighbours has long been used in Smooth Particle Hydrodynamics (SPH) and this technique can be applied to galaxies in simulations.  SPH calculates the density around a point by weighting each neighbour based on its distance from the point, with the smoothed galaxy density defined as
\begin{equation}
\rho=\sum_{i=1}^n W(|r_{i}|,h) ~.
\label{eq:sphdens}
\end{equation}
\noindent Here, $n$ is the number of neighbours used and $W(|r_{i}|,h)$ is the weighting given by
\begin{equation}
W(r,h)=\frac{8}{\pi h^3} \left\{
\begin{array}{ll}
1-6\left(\frac{r}{h}\right)^2 + 6\left(\frac{r}{h}\right)^3 &
0\le\frac{r}{h}\le\frac{1}{2} \\
2\left(1-\frac{r}{h}\right)^3 & \frac{1}{2}<\frac{r}{h}\le 1 ~, \\
0  & \frac{r}{h}>1 
\end{array}
\right.
\label{eq:kern}
\end{equation}
\noindent where $r$ is the distance to each neighbour and $h$ is the distance to the $n$-th nearest neighbour.  This weighting corresponds to the spline kernel of \citet{Monaghan85} and is the standard kernel of SPH\footnotemark.  This method was used with 20 neighbours in \citet{Park07}, but values of 32 and 64 are more common in SPH.

\footnotetext{We have adopted the notation of $h$ corresponding to the point at which the kernel equals zero as opposed to $2h$ as is used in traditional SPH literature. This is just a notational change.}

Another way to constrain local galaxy density using neighbours was proposed by \citet{Li11} for the Redshift One LDSS-3 Emission line Survey \citep[ROLES;][]{Gilbank10}.  \citet{Li11} considered the volume element of the nearest neighbour found by constructing a three dimensional cylinder using the five nearest neighbours to define its radius and depth.  In other words, this technique encloses the five nearest neighbours in a cylinder that no longer has to be centred on the galaxy being sampled, and leads to a better estimate of the relevant volume when compared with simply using a sphere of radius the fifth nearest neighbour.

Further consideration of the volume can be made by calculating the Voronoi volumes around each galaxy as a measure of the environment \citep[e.g.][]{Marinoni02, Cooper05}.  Voronoi volumes are polyhedrons constructed by bisecting the distance vectors to the nearest neighbours. Each galaxy will have a volume around it, for which it does not have to be at the centre, defining the points in space that are closer to it than any other galaxy. This gives an estimate of the local density.  Unlike the other neighbour-based methods, the number of neighbours used to define the shape of the volume probed is not fixed, which makes the technique fully adaptive.  For this study a projected Voronoi measurement is made by collapsing galaxies into two dimensional slices of $50\,h^{-1} {\rm Mpc}$ in depth.  The Voronoi shapes are then constructed on these surfaces to calculate the surface density of each galaxy.

\begin{figure*}
\psfrag{Halo Mass}[][][1][0]{$\rm log(M_{Halo}/M_{\odot})$}
\psfrag{Percentage Rank}[][][1][0]{Percentage Rank}
\includegraphics[width=160mm]{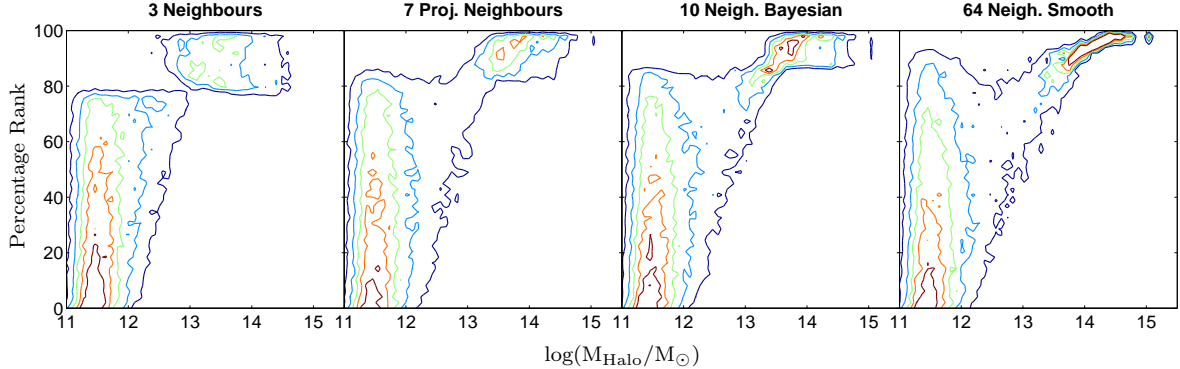}
\caption{The abundance of galaxies that have environments of a given percentage rank plotted against host halo mass, where environment is defined by the (from left to right) 3rd nearest neighbour, 7th projected nearest neighbour, 10 neighbour Bayesian metric and 64 neighbour kernel smoothed (SPH style).  Contours are linearly spaced showing regions of constant galaxy number.  The bimodal distribution is caused by the neighbour search remaining in or leaving the halo to find the next nearest neighbour.}
\label{fig:NHalo}
\end{figure*}

In Section~\ref{sec:res} we apply a number of the above nearest neighbour methods to the mock galaxy catalogue described in Section~\ref{sec:mod} and quantify their relative strengths, weaknesses and optimal applications.

\subsection{Fixed Aperture Environment Measures}
\label{sec:aperture}

In contrast to nearest neighbour methods, which define environment using a varying scale around each galaxy set by the distance to a pre-determined number of galaxy neighbours, fixed aperture methods instead probe a fixed area or volume around each galaxy, within which the number of neighbours are counted.  The more galaxies inside this area or volume, the denser the environment is assumed to be, and vice versa.  

Fixed aperture measures are often expressed as a density contrast, $\delta$, instead of a density, $\rho$.  Density contrast rescales the aperture count with respect to the mean and is typically defined as
\begin{equation}
\delta \equiv \frac{\delta \rho}{\rho} = \frac{N_g-\bar{N_g}}{\bar{N_g}} ~,
\end{equation}
\noindent where $N_g$ is the number of galaxies found in the aperture, and $\bar{N_g}$ is the mean number of galaxies that would be expected in the aperture if galaxies were instead distributed randomly throughout the entire volume.  

The fixed aperture technique was used by \citet{Croton05} to investigate the environments around galaxies in the 2dF Galaxy Redshift Survey \citep{Colless03}.  \citet{Croton05} used spherical apertures of radius $8\,h^{-1} {\rm Mpc}$, having investigated a range of sizes from of $4\,h^{-1} {\rm Mpc}$ to $12\,h^{-1} {\rm Mpc}$ \cite[see also][]{Abbas06}.

When distance information is not of sufficient accuracy (or absent), apertures in this methodology are instead projected on to the sky. Where possible, authors will then impose a velocity cut of order $\pm$1000 km $\rm s^{-1}$ to minimise interlopers \citep[e.g.][]{Gruetzbauch11}, for the same reasons discussed in Section \ref{sec:neigh}.  The magnitude of this velocity cut can vary depending on distance uncertainties.  This was investigated by \citet{Gallazzi09} who found velocity cuts of $\pm$6000 km $\rm s^{-1}$ ($dz=0.02$) represent the typical photometric redshift uncertainty and $\pm$500 km $\rm s^{-1}$ ($dz=0.0015$) represent the typical spectroscopic redshift uncertainty. Such errors can often have a detrimental effect on the measured density if not appropriately accounted for.  Note that when a velocity cut is imposed, an otherwise spherical aperture elongates into a cylinder in three dimensional space, within which galaxy counts are then taken. Whether this distortion is important for the environment measure depends on the focus of the analysis. Typical scales for the radius of an aperture range from $1\,h^{-1} {\rm Mpc}$ to $10\,h^{-1} {\rm Mpc}$, probing environments spanning individual haloes to large super-structures and voids in the cosmic web.

A variation on the fixed aperture method was proposed in \citet{Wilman10}, where counts were taken in annuli of increasing inner and outer radius, rather than within a single fixed aperture volume. This technique enables the larger scale environment to be probed and the influence of local regions around individual galaxies to be removed.  In its optimal form different sized annuli are applied in combination with apertures to better constrain the halo size and changes of environment with distance from the galaxy.

Finally, in addition to environment being defined by galaxy positions within the volume, we also measure environment as inferred from the background dark matter distribution. To obtain the neighbourhood dark matter environment in the Millennium Simulation the full volume is broken into a three dimensional grid with side-length $2\,h^{-1} {\rm Mpc}$. At the centre of each grid element a three dimensional Gaussian density is calculated using the local dark matter particles, smoothed over three different scales: $2.5$, $5$, and $10\,h^{-1} {\rm Mpc}$.  This Gaussian smoothed density is similar to the kernel smoothed density of Equation \ref{eq:sphdens}, but with a dark matter particle mass term in the sum.

In Section~\ref{sec:res} we apply a number of fixed aperture methods to the mock galaxy catalogue and measure local density around each galaxy. This allows us to quantify the properties that aperture measured densities best probe, and compare with the previously described nearest neighbour estimators.

\section{Results}
\label{sec:res}

To investigate the different properties of each galaxy environment measure, in this section we consider how they correlate with (1) the host dark matter halo mass, (2) the underlying dark matter environment, and (3) the colour of the galaxies. 

To facilitate this we have converted the output of each to a `percentage rank' for each galaxy.  This is computed by listing the galaxies in order of increasing density, then assigning them a percentage based on where they appear in that list, with zero percent being the least dense and one hundred percent the most dense.  Therefore, a galaxy with a percentage rank of ninety five has five percent of the galaxies in the sample denser than it and ninety five percent less dense than it.  This normalisation provides a fairer comparison between environment estimators and probes their relative rather than absolute distributions across the environment spectrum, which would otherwise be definition dependent.

\subsection{Dark Matter Halo Mass}

By design, the most fundamental property for a galaxy within our model is its dark matter halo mass.  Halo mass determines both the spatial distribution of the galaxy population and the individual galaxy properties.  Therefore, each environment measure should reveal some underlying correlation.  Typically halo masses of $\sim 10^{12}\,h^{-1} M_{\odot}$ correspond to the field, $\sim 10^{13.5}\,h^{-1} M_{\odot}$ to groups and $\sim 10^{15}\,h^{-1} M_{\odot}$ to clusters.

\subsubsection{Nearest neighbour results}

\begin{figure}
\psfrag{Halo Mass}[][][1][0]{$\rm log(M_{Halo}/M_{\odot})$}
\psfrag{Percentage Rank}[][][1][0]{Percentage Rank}
\includegraphics[width=82mm]{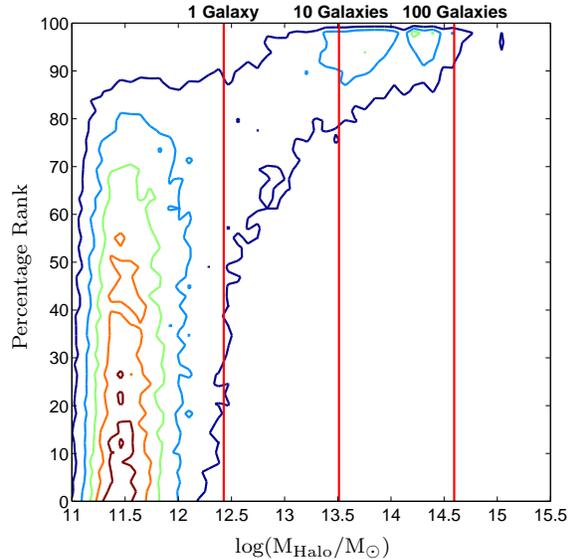}
\caption{The percentage rank of galaxy environments plotted against dark matter halo mass, as in Figure~\ref{fig:NHalo}, this time for the Voronoi method.  Contours are linearly spaced showing regions of constant galaxy number.  Vertical lines represent typical dark matter halo masses that host 1, 10 and 100 galaxies with $M_r-5{\rm log}(h)\leq-19$ (see Figure~\ref{fig:pop})}
\label{fig:vor}
\end{figure}

\begin{figure*}
\psfrag{Halo Mass}[][][1][0]{$\rm log(M_{Halo}/M_{\odot})$}
\psfrag{Percentage Rank}[][][1][0]{Percentage Rank}
\includegraphics[width=160mm]{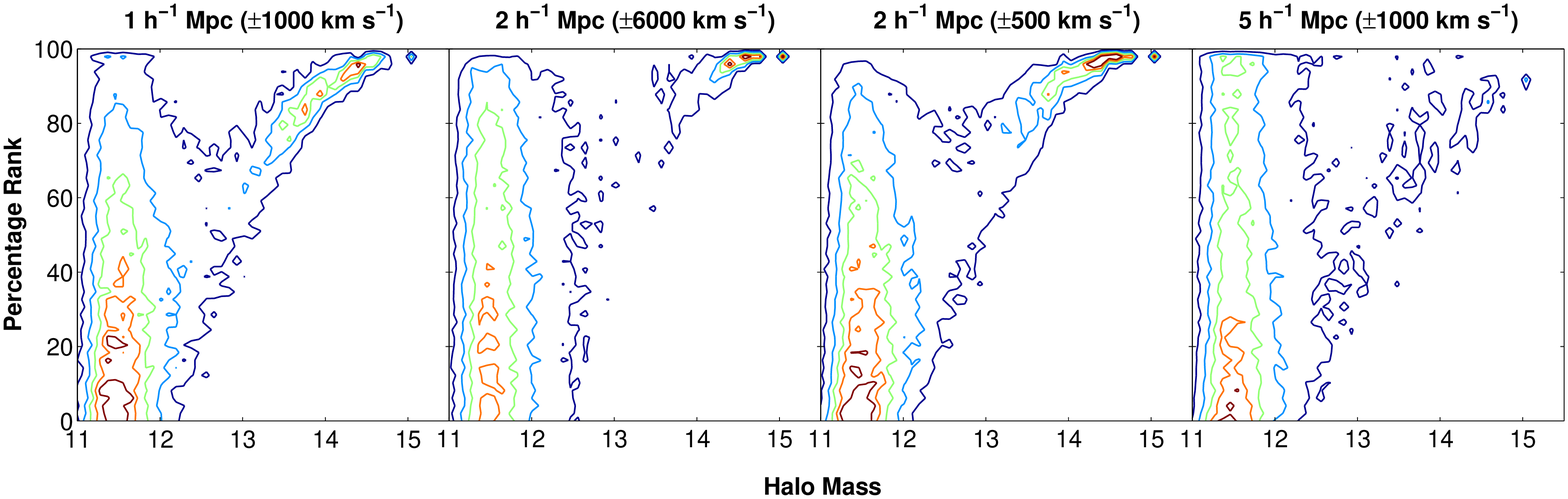}
\caption{The percentage rank of galaxy environment against dark matter halo mass, as in Figure~\ref{fig:NHalo}, for (from left to right) a $1\,h^{-1} {\rm Mpc}$ aperture with a velocity cut of $\pm1000\,\rm km\,s^{-1}$, a $2\,h^{-1} {\rm Mpc}$ aperture with a velocity cut of $\pm6000\, \rm km\,s^{-1}$, a $2\,h^{-1} {\rm Mpc}$ aperture with a velocity cut of $\pm500\, \rm km\,s^{-1}$ and a $5\,h^{-1} {\rm Mpc}$ aperture with a velocity cut of $\pm1000\, \rm km\,s^{-1}$. Contours are linearly spaced showing regions of constant galaxy number.}
\label{fig:AHalo}
\end{figure*}

Figure \ref{fig:NHalo} shows contours of the abundance of galaxies that have environments of a given percentage rank plotted against the host halo mass, for four different nearest neighbour-based techniques, with the number of neighbours increasing from left to right.  These are: the 3rd nearest neighbour density in three dimensions, the surface density for the projected 7th nearest neighbour, the three dimensional density using a 10 neighbour Bayesian metric, and the smooth kernel three dimensional density using 64 neighbours. 

The most noticeable feature of all panels in Figure~\ref{fig:NHalo} is that galaxies divide into two distinct groups, with the top $\sim 20$ percent dense environments occupied by galaxies in haloes more massive than $\sim 10^{12.5}\,h^{-1} M_{\odot}$, and the remaining $\sim 80$ percent of environments occupied by galaxies in haloes with masses lower than $\sim 10^{12.5}\,h^{-1} M_{\odot}$.  This bi-modality arises from the assumed association between galaxies and dark matter haloes required to fit the observed luminosity function and clustering observations, and is explored further below.

Looking in more detail, the lower $80$ percent of rank-ordered densities in Figure~\ref{fig:NHalo} shows no trend with halo mass, and as such, the term `local environment' no longer applies. In terms of a characteristic halo mass for a given environment, this result leaves individual galaxies near clusters indistinguishable from isolated galaxies in voids. 

In contrast, the behaviour of the high density--halo mass correlation depends on the neighbour method employed. In the highest $20$ percent environments, low $n$ neighbour searches smooth away any density dependence with halo mass. This can be seen by comparing the far left panel in Figure~\ref{fig:NHalo} (low $n$) with the far right panel (high $n$). As the number of neighbours used to define environment is increased, galaxies belonging to increasingly massive haloes (which host an increasing number of satellites) will be labelled as increasingly dense. Thus, to more precisely draw out the high density--halo mass environment correlations using nearest neighbour methods, a high $n$ is desirable.

The first two panels of Figure~\ref{fig:NHalo} provide an additional test of the importance of projection effects. Here, the 3rd nearest neighbour count is performed using three dimensional redshift space distances while the 7th nearest neighbour is performed with projected galaxy positions on the two dimensional sky.  Both methods show the same overall trend with halo mass. We find that, in general, projecting the galaxy positions simply blurs the edges of the two clouds with the overall shape preserved.

Another popular neighbour-based method used for measuring environment is Voronoi volumes, as discussed in Section~\ref{sec:neigh}.  Figure~\ref{fig:vor} shows how a Voronoi defined environment estimator also correlates with dark matter halo mass. We see a similar trend to that of the other neighbour-based methods, with the overall result close to the 7th nearest projected neighbour method shown in the second panel of Figure \ref{fig:NHalo}.

A comparison of Figure~\ref{fig:pop} with Figures~\ref{fig:NHalo} and \ref{fig:vor} reveals the origin of the bi-modality. Galaxies identified to be in the upper $20$ percent dense environments tend to be those whose neighbour search stays within the dark matter halo due to a large satellite population. Such haloes are almost always more massive than $\sim 10^{12.5}\,h^{-1} M_{\odot}$. In contrast, the lower $80$ percent density environments are identified by neighbour searches that extend beyond the halo due to a low or zero satellite population of significance. In general, haloes with few satellites almost always have masses smaller than $\sim 10^{12.5}\,h^{-1} M_{\odot}$, and neighbour searches will then tend to probe the inter-halo rather than inter-galaxy separations. 

\subsubsection{Fixed aperture results}

Many authors have employed fixed apertures to probe the local density around galaxies, as described in Section~\ref{sec:aperture}. In a similar vein to Figure~\ref{fig:NHalo}, Figure~\ref{fig:AHalo} shows how various aperture sizes correlate with host dark matter halo mass when a projected fixed aperture is employed with a cut in velocity space around each galaxy.  In addition, the central two panels show how the density--halo mass correlation changes if the velocity cut is increased for the same sized aperture.  This roughly corresponds to the difference one would expect with data having photometric vs. spectroscopic redshifts, as discussed in \citet{Gallazzi09}.  

The projected fixed aperture technique yields both similar and different trends when compared with the nearest neighbour technique shown in Figure~\ref{fig:NHalo}.  The overall shape is the same, with galaxies in haloes of mass less than $\sim 10^{12.5}\,h^{-1} M_{\odot}$ showing little correlation of halo mass with environment. At the high mass end there is a plume of increasing density that is much better defined than found with the nearest neighbour method (especially when compared to choices of low $n$).  This suggests that the fixed aperture methodology is a better probe of halo mass, especially for small apertures and velocity cuts.  There is however contamination at a fixed density from low mass haloes due to their close proximity to the high mass halo.

In particular, when there are enough galaxies to define the local large-scale structure, a fixed-scale environment probe is much more sensitive to the power-law nature of the two-point correlation function, where the abundance of close pairs falls off rapidly beyond the halo radius. This leads to the galaxy count in the fixed aperture also falling off rapidly. In contrast, nearest neighbour environment methods adapt the scale probed to keep signal-to-noise fixed. Hence, the division between a halo's interior and exterior becomes much less prominent.

At intermediate to low masses there is no relation between fixed aperture measured density and halo mass, and so the environment parameter breaks down, as is also the case for nearest neighbour environment parameters.  From an environment point-of-view, such haloes, which usually host galaxy groups, may be difficult to distinguish from cluster outskirts and from unassociated lower-mass haloes.

As the aperture is increased in size, the trend with halo mass fades when the aperture becomes much larger than the structures present.  For example, a super-cluster with a collective mass of $10^{16}\,h^{-1} M_{\odot}$ would have a radius\footnote{Radius here is determined by finding the scale at which the enclosed density is 200 times the critical density of the Universe.} of $\sim3.5 \,h^{-1} \rm Mpc$, smaller than the $5\,h^{-1} \rm Mpc$ aperture shown in the far right panel of Figure~\ref{fig:AHalo}.  When an aperture becomes large enough the contribution of individual haloes and structures blur and the environment--halo mass trend weakens or disappears. Hence, aperture size should be chosen carefully from the outset and be appropriate for the science questions of interest.

Finally, the two central panels of Figure \ref{fig:AHalo} illustrate the importance of velocity (or equivalently distance) uncertainties on the environment measure. Large velocity cuts, as is typically required with photometric data, make measuring environment with a fixed aperture ineffective.  This occurs for the same reason as using large apertures.  There, the aperture was wider than the structures of interest which smoothed out the signal, while here, the depth of the aperture scatters in superfluous counts from foreground and background objects, diluting any correlation. This does not apply to the highest mass clusters as they dominate the depth reducing the effect of interlopers.  Furthermore, any use of the angular correlation function as a probe of environment must first consider the redshift distribution of the galaxies and the uncertainties must be well understood \citep[e.g.][]{Coil04, Quadri08}.

\subsection{Galaxy Colour}

Galaxy colour has been shown to correlate with local galaxy density, with galaxies in over-dense environments being redder compared with those in under-dense environments (cf. cluster and field) \citep[e.g.][]{Lewis02, Kauffmann04, Cooper06, Gallazzi09}.  The model we employ in this paper has a constrained global $g-r$ colour distribution that mimics that of local galaxies in the SDSS \citep{Skibba09b}. Hence, the degree to which different environment metrics can recover this relation can be tested.

\begin{figure*}
\psfrag{g-r}[][][1][0]{$g-r$}
\psfrag{fraction}[][][1][0]{fraction}
\includegraphics[width=160mm]{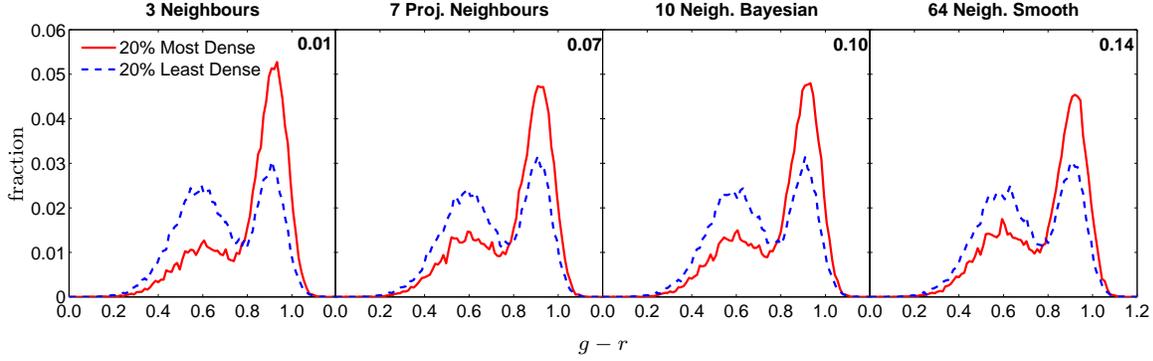}
\caption{Histograms of galaxy colour for the 20 percent most dense (red solid) and 20 percent least dense (blue dashed) galaxies, measured using $n$-th nearest neighbour statistics, defined by the (from left to right) 3rd nearest neighbour, 7th projected nearest neighbour, 10 neighbour Bayesian metric and 64 neighbour kernel smoothed (SPH style). The number in the upper right of each panel is the Kolmogorov-Smirnov probability that both samples are drawn from the same distribution.}
\label{fig:NCol}
\end{figure*}

Figure \ref{fig:NCol} shows histograms of the $g-r$ colour distribution for the 20 percent most dense and 20 percent least dense galaxies defined with the same four nearest neighbour methods used in Figure~\ref{fig:NHalo}: the 3rd nearest neighbour density in three dimensions, the projected 7th nearest neighbour, density defined from a 10 neighbour Bayesian metric, and the smooth kernel density using 64 neighbours.  In the $20$ percent most dense environments, all nearest neighbour-based environment measures show a clear red peak and a more weakly populated blue cloud. In contrast, in the lowest $20$ percent of environments galaxies are split more evenly between the red and blue populations. As the neighbour number is increased (from left to right), there are only small changes in the relative colour distributions in environment extremes. The Kolmogorov-Smirnov probability that both samples are drawn from the same distribution is shown in the upper right of each panel.

\begin{figure*}
\psfrag{g-r}[][][1][0]{$g-r$}
\psfrag{fraction}[][][1][0]{fraction}
\includegraphics[width=160mm]{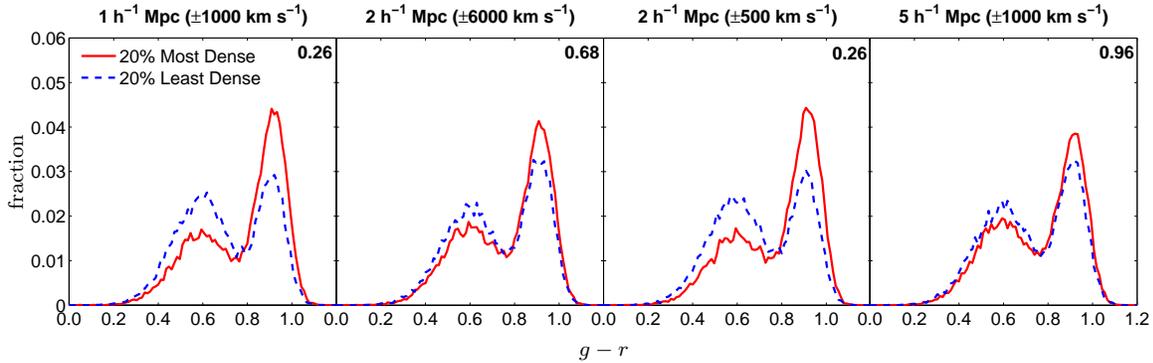}
\caption{Histograms of galaxy colour, as in Figure~\ref{fig:NCol}, for the 20 percent most dense (red solid) and 20 percent least dense (blue dashed) galaxies, measured using (from left to right) a $1\,h^{-1} {\rm Mpc}$ aperture with a velocity cut of $\pm1000\,\rm km\,s^{-1}$, a $2\,h^{-1} {\rm Mpc}$ aperture with a velocity cut of $\pm6000\, \rm km\,s^{-1}$, a $2\,h^{-1} {\rm Mpc}$ aperture with a velocity cut of $\pm500\, \rm km\,s^{-1}$ and a $5\,h^{-1} {\rm Mpc}$ aperture with a velocity cut of $\pm1000\, \rm km\,s^{-1}$. The number in the upper right of each panel is the Kolmogorov-Smirnov probability that both samples are drawn from the same distribution.}
\label{fig:ACol}
\end{figure*}

\begin{figure*}
\psfrag{Dark Matter Percentage Rank}[][][1][0]{Dark Matter Percentage Rank}
\psfrag{Galaxy Percentage Rank}[][][1][0]{$\delta_8$ Galaxy Percentage Rank}
\includegraphics[width=140mm]{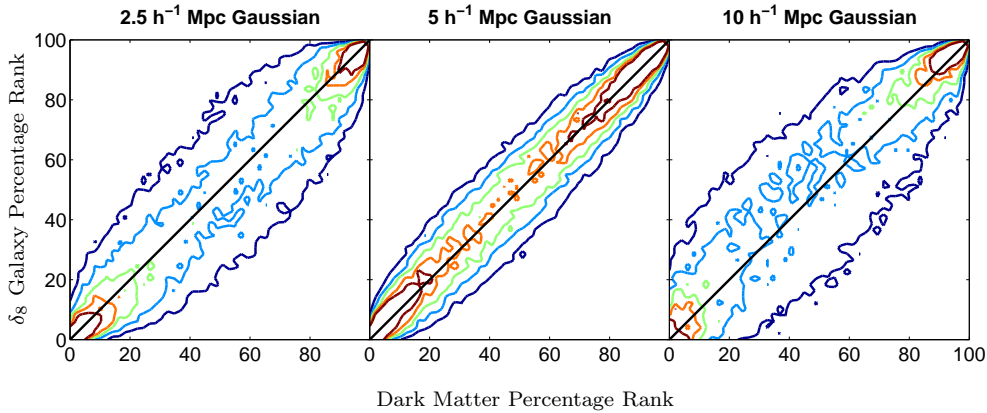}
\caption{The percentage rank of central galaxy environment using an $8\,h^{-1} {\rm Mpc}$ spherical aperture plotted against the percentage rank of background dark matter environment measured using a smooth Gaussian filter of radius (from left to right) $2.5\,h^{-1} {\rm Mpc}$, $5\,h^{-1} {\rm Mpc}$ and $10\,h^{-1} {\rm Mpc}$.  Contours are linearly spaced showing regions of constant galaxy number.}
\label{fig:dme}
\end{figure*}

Figure~\ref{fig:ACol} shows histograms of colour for the 20 percent most dense and 20 percent least dense galaxies as probed by fixed apertures of various size, as used previously in Figure~\ref{fig:AHalo}.  The central two panels show how these distributions change if the velocity cut is increased or decreased for the same sized aperture.  This roughly corresponds to the difference between photometric and spectroscopic redshift uncertainties \citep{Gallazzi09} (see Section~\ref{sec:aperture}). For small apertures, the colour distributions of both density extremes look remarkably similar to that found for the nearest neighbour methods shown in Figure~\ref{fig:NCol}. However, as the volume of the fixed aperture is increased similar trends to that found in the previous section emerge. In particular, as the aperture becomes larger (either in radius or depth), the differences between the colour distributions of galaxies in environment extremes lessen. Here, the individual properties of galaxies are smoothed over due to the large variety of local environments falling within the aperture. For apertures probing scales much larger than the typical cluster the distinction between environments vanishes. This suggests that environment questions relating to galaxy colour (or properties that correlate with colour) should avoid fixed aperture methods with large smoothing radii or depth \citep[e.g.][]{Croton05}.  Furthermore, the Kolmogorov-Smirnov probabilities indicate that nearest-neighbour based methods detect stronger colour-environment relations than all the apertures tested here.

\subsection{Dark Matter Environment}

Dark matter haloes are known to be biased tracers of the underlying dark matter distribution, and it is interesting to compare how haloes and the smooth background mass field correlate with respect to their environment ranking, and how this relates to the galaxy distribution.  To this end, the simulation volume has been divided using a three dimensional grid of side-length $2\,h^{-1} {\rm Mpc}$, and the neighbourhood dark matter density field measured with a Gaussian filter placed at the centre of each grid element, smoothed on three different scales: $2.5$, $5.0$ and $10\,h^{-1} {\rm Mpc}$ (see Section~\ref{sec:aperture}). We compare this to the environment measured directly from central galaxy counts within a fixed spherical aperture of radius $8\,h^{-1} {\rm Mpc}$ \citep{Croton05}.

Figure~\ref{fig:dme} shows how the background dark matter density, Gaussian smoothed on various scales, correlates with the large-scale galaxy density, top-hat smoothed on an $8\,h^{-1} {\rm Mpc}$ scale. The correlation is weakest for the smallest Gaussian smoothing scale of $2.5\,h^{-1} {\rm Mpc}$, becomes tighter at a scale of $5\,h^{-1} {\rm Mpc}$, before becoming weaker again at $10\,h^{-1} {\rm Mpc}$. The point of tightest correlation between dark matter and galaxy measured density approximately corresponds to the same physical scale being probed by each in three--dimensional space. At fixed dark matter density the scatter in density measured by galaxies is approximately $40$ percent. This indicates the degree of precision with which one can probe the smooth background density using galaxies as tracers of the mass distribution.

We have compared the other environment measures used in this paper to the background dark matter density but omit the figures for brevity. In short, a similar trend to Figure~\ref{fig:dme} is found for the 64 neighbour smooth density environment measure, but with the tightest correlation at a radius of $2.5\,h^{-1} {\rm Mpc}$.  For the other neighbour and small aperture methods, weak correlations are found when plotted against a dark matter density smoothing scale of $2.5\,h^{-1} {\rm Mpc}$ but which disappear on larger scales.  Environments measured in annuli and projected aperture methods that impose a photometric-type redshift velocity cut show no correlation on any scale due to only the largest clusters dominating the depth cut, while the 10 neighbour Bayesian metric and 20 neighbour smooth density again show a similar correlations to the dark matter smoothing scale of $2.5\,h^{-1} {\rm Mpc}$.

\subsection{Individual Galaxies}
\label{sec:ind}

\begin{figure}
\includegraphics[width=82mm]{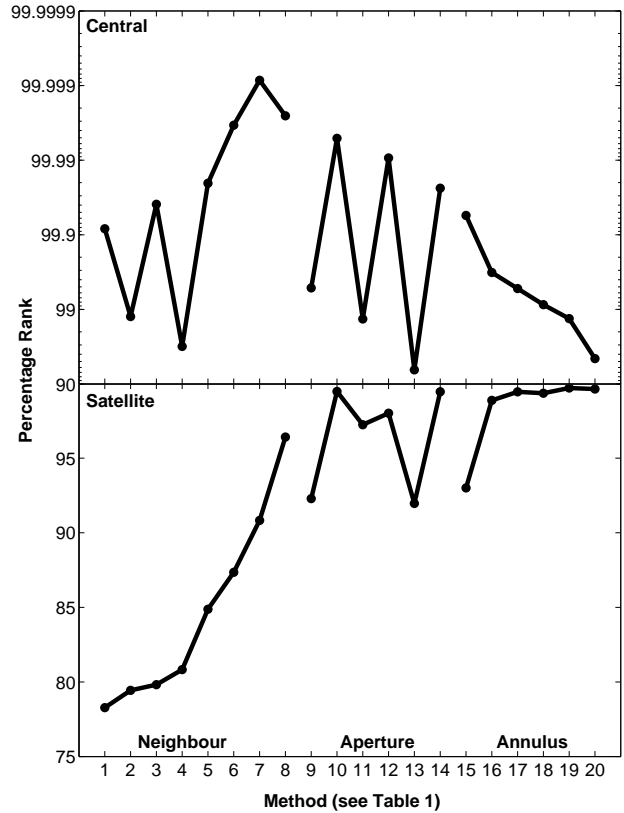}
\caption{(top) The percentage rank of various density estimators (see Table~\ref{tab:comp}) for a single central galaxy living in the fourth most massive halo in the simulation. The density methods are grouped by increasing neighbour number, increasing aperture and increasing inner radius of an annulus. (bottom) The same as the top panel, but this time the percentage density ranking of the outer most satellite galaxy in the same halo for each method.}
\label{fig:comp}
\end{figure}

In the previous sections we investigated how different environment parameters correlate with different galaxy properties in a statistical sense by considering the whole sample.  As implied by Figures~\ref{fig:NHalo} and \ref{fig:AHalo}, when selecting the most and least dense environments different methods will potentially select different galaxy populations. An alternative and complementary way to compare the different environment methodologies is to consider individual galaxies in the mock catalogue and examine how each measure ranks them relative to the others. By considering individual galaxies a better understanding of why these galaxies were chosen can be obtained. This also highlights the consistency (or lack thereof) between different definitions of environment. Below we present one example that is representative of the general behaviour for high mass haloes.

The top panel of Figure~\ref{fig:comp} shows how the different environment measures listed in Table~\ref{tab:comp} compare when one focuses on the central galaxy occupying the fourth most massive halo in the simulation, with mass $10^{15.08}\,h^{-1} M_{\odot}$. The environment measures are separated into three groups based on the technique they use: neighbours, aperture and annulus. All environment measures place this galaxy within the top 10 percent of rank ordered densities in the simulation volume, with the majority placing it within the top 1 percent.  When considering annuli to define environment, the top panel of Figure~\ref{fig:comp} shows that the further one moves from the centre of the halo the lower the rank density measured.  This simply highlights that the outer regions of a halo tend to be less dense than the core.  When considering aperture methods there is less of a trend between different definitions. However, for a fixed depth, increasing the aperture size reduces the rank density measured, while for fixed aperture size, the density rank appears sensitive to the inclusion of both the halo core (smaller velocity cut) and full extent (larger velocity cut).  As mentioned in previous sections, the larger velocity cut used to represent photometric redshift uncertainties has a smaller effect on large clusters as the cluster members dominate the galaxies within the depth cut.  For neighbour-based methods there is a general increase in the rank density as the neighbour number increases.  This is due to the increased neighbour count contributing from within the galaxy halo. Specifically, as the number of neighbours increases galaxies in smaller haloes are demoted down the rank list, and so the galaxies in large haloes are promoted.

The bottom panel of Figure \ref{fig:comp} shows how the different density estimators rank the most distant satellite associated with the central galaxy of the same $10^{15.08} \rm M_{\odot}$ halo used in the top panel. This is a test of how satellites on the outskirts of cluster environments would be classified in each density scheme. The range of environment ranking is much larger between the different measures, with the apertures and annuli mostly finding higher rank densities than the neighbour estimates.  This again comes back to scale, with neighbour methods probing the internal properties of the outer halo, and aperture and annulus methods being sensitive to the larger structure of the halo and its surrounds.  Additionally, the trend of increasing rank density with increasing neighbour number is again seen as the neighbour count reaches deeper into the halo core from the boundary.

\section{Discussion and Summary}
\label{sec:discussion}

The phrase ``galaxy environment'' is a very general concept that has been used in the literature in a variety of ways.  Its definition -- what it measures and how it is measured -- can vary from author to author.  This creates uncertainty when trying to compare results for environmental trends.  In practise, galaxy environment is quantified in one of two ways: by the distance to the $n$-th nearest neighbour or by using a fixed aperture to probe the surrounds.  Over the course of time these two methods have evolved in the literature.  However, both methods and their variants provide a measure of the density field surrounding a galaxy and hence can be used to answer specific environment-related questions.

To fairly compare many different environment measures one would ideally like to use a common galaxy catalogue as a starting point.  This was achieved in the present work by applying a halo occupation distribution model to the $z=0$ output of the Millennium dark matter simulation.  Our model is designed to accurately reproduce the luminosity, colour and spatial distribution of galaxies in the Sloan Digital Sky Survey.  The resulting data cube was also used to generate a mock light cone so that the environment measures could be applied in a more realistic geometry.

The major conclusions of this paper are as follows:

\begin{itemize}

\item \textbf{Galaxy environment versus halo mass}: Comparing neighbour and aperture based environment measures to the dark matter halo mass of a galaxy reveals how \textit{they measure different aspects of the halo}. In particular, nearest neighbour methods that use a small enough neighbour number best probe the internal properties of the halo.  For haloes that contain fewer galaxies than the neighbour number, the inter-halo separation dominates the calculation and galaxy--environment correlations tend to wash out.  In contrast, \textit{aperture measures tend to better probe the halo as a whole} and so lead to larger density values corresponding to larger haloes, which more accurately reflect their larger masses.  A smaller aperture than those studied here could be used to probe cluster environments on a scale similar to the nearest neighbour-based methods, but these would be unsuitable for the field due to the distance between neighbours being too large.  This is in agreement with the findings of \citet{Haas11}.

\item \textbf{Galaxy environment within haloes}: \textit{Galaxy density internal to a halo's boundary is found to be independent of its mass when probed using the neighbour method}.  While galaxies at the edge of a halo are always in less dense environments than those at the centre, the galaxy environment at the centre of intermediate mass haloes is approximately the same as that at the centre of very massive ones.\footnote{The concentration and mass of dark matter haloes are anti-correlated, and since the number density distribution of galaxies follows that of the dark matter particles \citep{Yang05}, the central concentration of galaxies should also vary slightly with halo mass.  In practice, however, the trend is difficult to detect observationally.}  By fitting the number of galaxies for a given halo mass, we find that the number of galaxies per unit dark matter mass is constant and this is in agreement with the findings of \citet{Poggianti10}.

\item \textbf{Environmental dependence of galaxy colour}: When comparing how the different environment measures distribute galaxy colour, almost all methods recover the observed correlation that galaxies are redder in denser environments compared to those in less dense environments.  \textit{This relation only disappears for very large apertures, of order $>5\,h^{-1} {\rm Mpc}$}. On scales larger than this the most dense and least dense galaxies are found to have similar colour distributions.  Here, the aperture is large enough to encompass a statistically representative number of different haloes, resulting in a smoothing out of the colour differences over such large volumes. \textit{This behaviour is also expected to extend to any property that correlates strongly with colour}.

\item \textbf{Galaxy environment versus dark matter environment}: On the other hand, \textit{very large fixed apertures are the most accurate at recovering the large-scale dark matter environment}. For example, an $8\,h^{-1} {\rm Mpc}$ spherical aperture used to calculate the galaxy density correlates well with the dark matter environment measured using Gaussian smoothing on $5\,h^{-1} {\rm Mpc}$ scales. Similar results are found with high number $n$-th neighbour estimates. The important parameter here is scale, with larger probed scales better correlated than smaller scales.

\item \textbf{Environment measures using spectroscopic versus photometric redshifts}: In addition to the environment measures themselves, we also explored the general effects of photometric and spectroscopic redshift uncertainties by varying the velocity cut used to calculate projected environment. For a typical photometric redshift uncertainty most trends with environment disappear or become significantly weaker. This is caused by the depth of the aperture becoming much larger than the objects being probed, and the scattering of interlopers which contaminate the density probe. This effect decreases for the largest clusters as the members dominate the depth cut.  We warn that \textit{photometric redshifts may be unsuitable for measuring certain properties when using a range of environmental scales}, especially at high redshifts.

\item \textbf{Environments of central and satellite galaxies}: On a galaxy-by-galaxy basis, most environment methods agree on the relative environment rank of central galaxies in massive haloes (to within a few percent). \textit{There is less agreement with the satellite population in clusters}, where the result becomes more sensitive to the method employed.

\end{itemize}

Dark matter haloes are often broadly categorised as residing in `field', `group', and `cluster' environments based solely on their mass. In Figure~\ref{fig:pop}, for example, haloes of mass $M\sim10^{12}M_\odot$, $10^{13.5}M_\odot$, and $10^{15}M_\odot$ approximately correspond to these environment bins, respectively.  Many environment analyses use this categorisation, although as we have seen, the distinctions between them can often be blurred in detail.  Some studies also attempt to explicitly identify galaxies that are isolated or in groups or rich clusters, for example using friends-of-friends group-finding algorithms \citep[e.g.][]{Berlind06, Yang07}.  Analyses using group catalogues are complementary to studies with nearest neighbour or fixed aperture measures, or with galaxy clustering \citep[e.g.][]{Weinmann06, Martinez06, Blanton07, Bosch08, Balogh09, Skibba11}.  Work focused on galaxy clusters have also yielded complementary results \citep[e.g.][]{Poggianti08, Rudnick09, Bamford09, Wolf09, Gallazzi09}. 

Importantly, the way a galaxy forms and evolves is clearly related to its environment. Some galaxy properties, such as luminosity, colour, and stellar mass, are directly correlated with the large-scale environment through the host dark matter halo \citep[e.g.][]{Zehavi05, Li06}.  Other galaxy properties, such as structure, are to some extent only indirectly correlated with the environment \citep[e.g.][]{Kauffmann04, Blanton05, Cassata07}.  Indeed, a number of authors report that, for many aspects of the galaxy population, environmental dependence is often weak once stellar mass is fixed \citep{Bosch08, Bosch08b, Scodeggio09, Bolzonella10, Vulcani11}.  In any case, these studies highlight the fact that it is important to carefully determine how a galaxy's environment is characterised, and to identify and navigate the potential aspects of the environment analysis that may bias the results.

The key consideration when picking an environment measure is the scale that is being probed.  The term environment is very general but in fact breaks down into two main regions, and we argue that the community should agree on a standard terminology for clarity and to avoid future confusion.  The first region is the `local environment' which corresponds to scales internal to a halo.  These are best probed using nearest neighbour methods, but the value of $n$ is important.  When $n$ is larger than the number of galaxies likely to reside within the halo the usefulness of this environment measure can weaken.  The second region lies external to the halo, the `large-scale environment'.  The large-scale environment is best probed using aperture based methods. In general, there is no simple way to probe all environments with a single method, and one should consider carefully the best tool to answer the questions at hand.

This paper marks the first in a series exploring the meaning and methods of galaxy environment, as measured in the current literature.  In the present work we have focused on using a clean sample of mock galaxies to quantify how selected properties of the galaxy population correlate with different environment methods, and how these methods themselves compare. Future work will include investigating the detrimental effects of survey geometry, edges and holes (such as those caused by stars) on environment and techniques that can be applied to successfully overcome them. Furthermore, the relationship between galaxy, halo and dark matter environment warrants additional exploration, as does the redshift dependence of a galaxy's environment (defined in various ways), what the different environment methods tell us about galaxy evolution, and how these can best be applied to the noisy data of the high redshift Universe.

\section*{Acknowledgments}

The authors wish to thank Hanni Lux and Idit Zehavi for useful discussions and the referee, Christian Wolf, for improving the quality of the paper.

DJC acknowledges receipt of a QEII Fellowship by the Australian Research Council.  HBA thanks NRF for the research grant 2010-0023319.  NBC is supported by a CIERA Postdoctoral Fellowship.  MEG is supported by an STFC Advanced Fellowship.  The Dark Cosmology Centre is funded by the Danish National Research Foundation.

The Millennium Simulation used in this paper was carried out by the Virgo Supercomputing Consortium at the Computing Centre of the Max-Planck Society in Garching.  The halo merger trees used in the paper are publicly available through the GAVO interface, found at http://www.mpa-garching.mpg.de/millennium/.

The mock galaxy catalogue generated from the Millennium Simulation trees can be downloaded from the Theoretical Astrophysical Observatory (TAO) repository: http://tao.it.swin.edu.au/mock-galaxy-factory/.

\bsp

\label{lastpage}

\end{document}